\documentclass[aps,prl,twocolumn,showpacs,superscriptaddress]{revtex4-1}

\usepackage {graphicx}
\usepackage{epsf}

\begin{document}

\title{Simultaneous Measurements of the Beta Neutrino Angular Correlation in $^{32}$Ar Pure Fermi and Pure Gamow-Teller Transitions using Beta-Proton Coincidences}

\author{V. Araujo-Escalona}
\affiliation{Instituut voor Kern- en Stralingsfysica, Katholieke Universiteit Leuven, B-3001 Leuven, Belgium}
\author{D. Atanasov}
\altaffiliation[Present address:]{ CERN, Geneva, Switzerland}
\affiliation{Instituut voor Kern- en Stralingsfysica, Katholieke Universiteit Leuven, B-3001 Leuven, Belgium}
\author{X. Fl\'echard}
\email{flechard@lpccaen.in2p3.fr}
\affiliation{Normandie Univ, ENSICAEN, UNICAEN, CNRS/IN2P3, LPC Caen, 14000 Caen, France}
\author{P. Alfaurt}
\affiliation{CENBG, UMR5797, Universit\'e de Bordeaux, CNRS, F-33170 Gradignan, France}
\author{P. Ascher}
\affiliation{CENBG, UMR5797, Universit\'e de Bordeaux, CNRS, F-33170 Gradignan, France}
\author{B. Blank}
\affiliation{CENBG, UMR5797, Universit\'e de Bordeaux, CNRS, F-33170 Gradignan, France}
\author{L. Daudin}
\affiliation{CENBG, UMR5797, Universit\'e de Bordeaux, CNRS, F-33170 Gradignan, France}
\author{M. Gerbaux}
\affiliation{CENBG, UMR5797, Universit\'e de Bordeaux, CNRS, F-33170 Gradignan, France}
\author{J. Giovinazzo}
\affiliation{CENBG, UMR5797, Universit\'e de Bordeaux, CNRS, F-33170 Gradignan, France}
\author{S. Gr\'evy}
\affiliation{CENBG, UMR5797, Universit\'e de Bordeaux, CNRS, F-33170 Gradignan, France}
\author{T. Kurtukian-Nieto}
\affiliation{CENBG, UMR5797, Universit\'e de Bordeaux, CNRS, F-33170 Gradignan, France}
\author{E. Li\'enard}
\affiliation{Normandie Univ, ENSICAEN, UNICAEN, CNRS/IN2P3, LPC Caen, 14000 Caen, France}
\author{G. Qu\'em\'ener}
\affiliation{Normandie Univ, ENSICAEN, UNICAEN, CNRS/IN2P3, LPC Caen, 14000 Caen, France}
\author{N. Severijns}
\affiliation{Instituut voor Kern- en Stralingsfysica, Katholieke Universiteit Leuven, B-3001 Leuven, Belgium}
\author{S. Vanlangendonck}
\affiliation{Instituut voor Kern- en Stralingsfysica, Katholieke Universiteit Leuven, B-3001 Leuven, Belgium}
\author{M. Versteegen}
\affiliation{CENBG, UMR5797, Universit\'e de Bordeaux, CNRS, F-33170 Gradignan, France}
\author{D. Z\'akouck\'y}
\affiliation{Nuclear Physics Institute, Acad. Sci. Czech Rep., CZ-25068 Rez, Czech Republic }
\date{\today}

\begin{abstract}
We report first measurements of the beta-neutrino angular correlation based on the kinetic energy shift of protons emitted in parallel or anti-parallel directions with respect to the positron in the beta decay of $^{32}$Ar.
This proof of principle experiment performed at ISOLDE/CERN provided simultaneous measurements for both a superallowed 0$^+$~$\rightarrow$~0$^+$ transition and a Gamow-Teller transition followed by proton emission.
The results, respectively ${\tilde a_{\beta\nu}}=1.007(32)_{stat}(25)_{syst}$ and ${\tilde a_{\beta\nu}}=-0.222(86)_{stat}(16)_{syst}$, are found in agreement with the Standard Model. 
The analysis of the data shows that future measurements can reach a precision level of 10$^{-3}$ for both pure Fermi and pure Gamow-Teller decay channels, providing new constraints on exotic weak interactions.
\end{abstract}

\maketitle
\section{\label{sec1}INTRODUCTION}
Precision measurements in nuclear and neutron beta decays are competitive tools to search for new physics and perform symmetry tests of the standard model (SM) in the electroweak sector. For a collection of selected transitions, they provide constraints that are complementary to high energy physics experiments \cite{Gonzalez19}. In particular, the beta-neutrino angular correlation coefficient ${a_{\beta\nu}}$ gives direct access to possible contributions of scalar $(S)$ or tensor $(T)$ couplings, involving other bosons than the $W^{\pm}$ ones associated to the standard vector $(V)-$axial-vector $(A)$ couplings of the weak interaction.
Assuming maximal parity violation and no time-reversal symmetry violation for the standard $V-A$ components of the interaction, the angular correlation coefficient ${a_{\beta\nu}}$ can be expressed as 
\begin{eqnarray}
a_{\beta\nu_{F}} \approx 1-\frac{ {\vert}C_{S}{\vert}^2 + {\vert}C^{'}_{S}{\vert}^2 } {{\vert}C_{V} {\vert}^2}\label{a_F}\\
a_{\beta\nu_{GT}} \approx -\frac{1}{3} \Big[1-\frac{ {\vert}C_{T}{\vert}^2 + {\vert}C^{'}_{T}{\vert}^2 } {{\vert}C_{A} {\vert}^2}\Big]\label{a_GT}
\end{eqnarray}
for pure Fermi (F) transitions and for pure Gamow-Teller (GT) transitions, respectively, where $C_i$ and $C_{i}^{'}$, ($i=V,A,S,T$) are the fundamental weak coupling constants.
The beta-neutrino angular correlation is accessible through the decay rate of unpolarized nuclei \cite{Jackson57}
\begin{eqnarray}
w(E_e,\Omega_e,\Omega_{\nu}) ~\propto~ w_0(Z,E_e)\Big(1+\frac{\vec{p_e} \cdot \vec{p_\nu}} {E_e E_\nu}a_{\beta\nu}+\frac{m_e}{E_e}b\Big)~~\label{w}
\end{eqnarray}
where $w_0(Z,E_e)$ includes the phase space factor and the Fermi function, $p_{e,\nu}$ and $E_{e,\nu}$ are the momenta and energies of the beta particle $(e)$ and of the neutrino $(\nu)$, $m_e$ is the rest mass of the electron and $b$ is the Fierz interference term. For pure F and pure GT transitions, respectively, this Fierz term is given by
\begin{eqnarray}
b_{F} \approx \pm Re\Big(\frac {C_{S}+C^{'}_{S}}{C_{V}}\Big) ~~~~~ b_{GT} \approx \pm Re\Big(\frac {C_{T}+C^{'}_{T}}{C_{A}}\Big)\label{b_F}
\end{eqnarray}
($\pm$ sign referring to $\beta^{\pm}$ decays).
As it is difficult to measure independently both ${a_{\beta\nu}}$ and $b$ in Eq.~(3), the observable extracted from most experiments is $\tilde{a}_{\beta\nu}\approx{a_{\beta\nu}}/(1+\alpha b)$, where the coefficient $\alpha$ gives the sensitivity to the Fierz term and can be determined by means of simulations \cite{Gonzalez16}.
The constraints set on exotic couplings originate then from the dependence of $\tilde{a}_{\beta\nu}$ on both ${a_{\beta\nu}}$ and $b$.
Today, nuclear and neutron beta decay limits on $S$ and $T$ couplings (relative to $V$ and $A$ couplings) involving either right-handed (${C_{S,T}=-C^{'}_{S,T}}$) or left-handed (${C_{S,T}=C^{'}_{S,T}}$) neutrinos are at the 10$^{-2}$ resp. 10$^{-3}$ level, requiring experimental precisions at the 10$^{-3}$ level \cite{Gonzalez19}.

Due to momentum and energy conservation laws, the angular correlation between the two leptons impacts the momentum distribution of the recoiling nucleus. The value of ${\tilde a_{\beta\nu}}$ can thus be inferred either from a direct measurement of the daughter nucleus recoil energy \cite{Johnson63, Gorelov05, Vetter08, Flechard11, Finlay16} or by observing secondary particles emitted after the decay \cite{Clifford89,Egorov97,Adelberger99,Vorobel03,Sternberg15}.
Both techniques yield similar constraints on exotic couplings with ${\Delta\tilde a_{\beta\nu}}\sim5.10^{-3}$ for pure F transitions \cite{Gorelov05, Adelberger99}, and ${\Delta\tilde a_{\beta\nu}}\sim3.10^{-3}$ for pure GT ones \cite{Johnson63, Sternberg15}.
Ongoing experimental programs aim today at precision levels of $10^{-3}$ and below \cite{Behr14, Leredde15, Shidling19, Blank16, Severijns17, Mukul17, Burkey19}.

The present project targets a similar goal by improving by a factor 5 or more the most precise results previously obtained from the recoil energy broadening of beta-delayed protons in the decay of $^{32}$Ar towards its isobaric analogue state in $^{32}$Cl \cite{Adelberger99, Garcia00, Blaum03}. 
Fig.~1 shows the simplified decay scheme of $^{32}$Ar where the 0$^+$~$\rightarrow$~0$^+$ and the pure Gamow-Teller transitions of interest are indicated.
Instead of a broadening, the present experiment, called WISArD (Weak Interaction Studies with $^{32}$Ar Decay), measures the kinetic energy shift of protons emitted in parallel or anti-parallel directions with respect to the positron. This beta-proton coincidence technique drastically reduces the influence of the proton detector response function and of the intrinsic proton peak shape. 
It also increases the statistical sensitivity on $\tilde{a}_{\beta\nu}$: Monte Carlo simulations of the experiment with the present setup show that the statistical uncertainty on $\tilde{a}_{\beta\nu}$ is reduced by a factor $\sim2.5$ when using the proton peak energy shift technique instead of the peak broadening technique. The effective gain in sensitivity should be even higher as this factor was obtained assuming a perfectly known proton detector response function. In real experiments, the uncertainty on the detector response function would affect significantly the precision for a broadening measurement, but not for a shift measurement.
Moreover, this new technique allows simultaneous measurements with beta-delayed protons resulting from both pure F and pure GT transitions of the $^{32}$Ar nucleus (Fig.~1).
Note that a similar approach is currently undertaken by the TAMUTRAP experiment \cite{Shidling19} using a Penning trap to confine radioactive ions.
\begin{figure}[h!]
\includegraphics[width=\columnwidth]{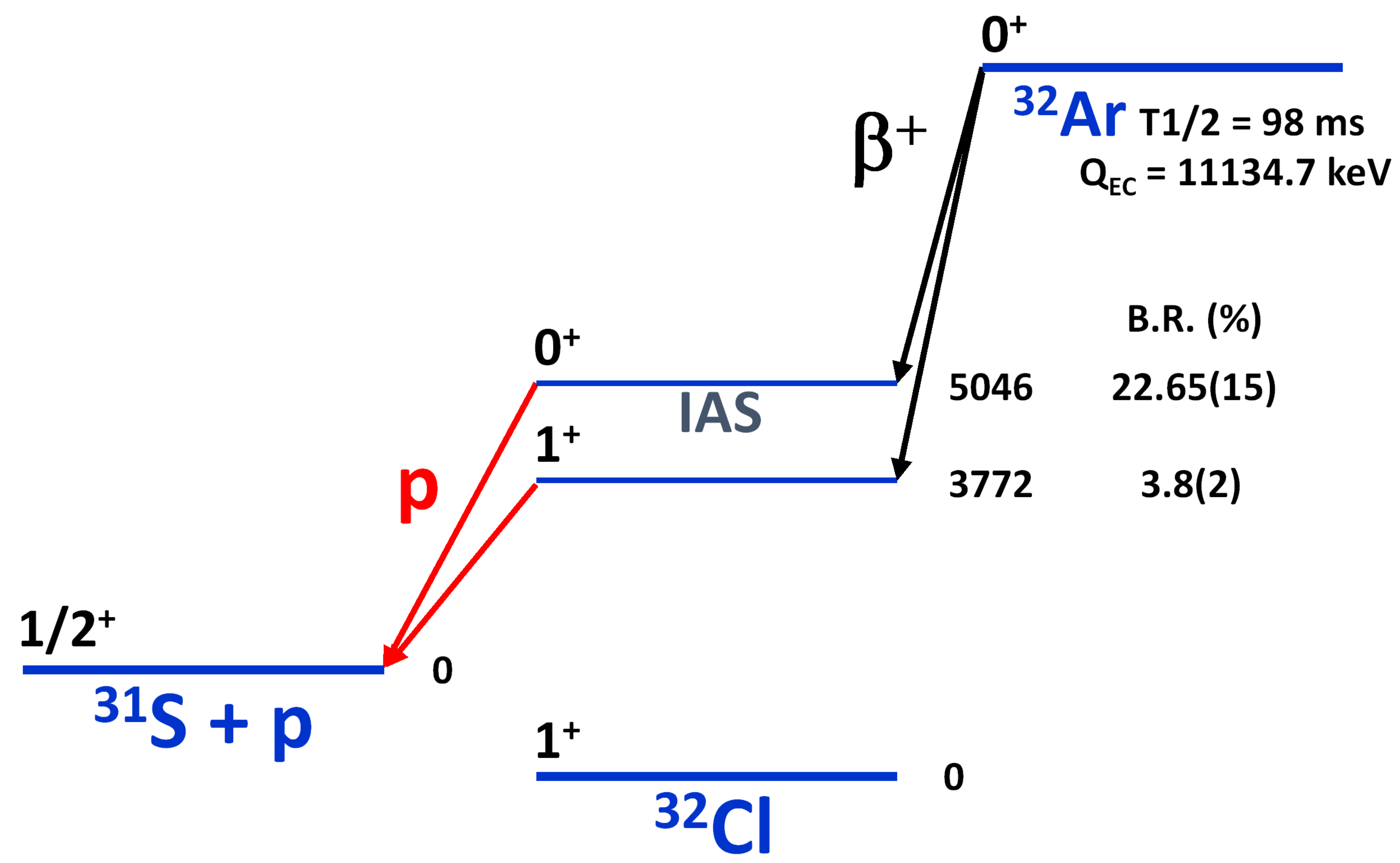}
\caption{(Color online) Simplified $^{32}$Ar decay scheme. Only relevant transitions discussed in the present paper are indicated.}
\label{fig1}
\end{figure}

\section{\label{sec2}EXPERIMENT}
While a dedicated set-up for WISArD is still under development, a proof of principle experiment was performed at ISOLDE-CERN with equipment and detectors readily available and the details of which will be published separately \cite{Atanasov}.
The detection setup, shown in Fig.~2, is installed in the vertical superconducting solenoid of the former WITCH experiment \cite{Finlay16}. It comprises eight $300~\mu m$ thick silicon detectors with effective diameter $\phi=30$~mm for protons and a $\phi=20$~mm, $L=50$~mm plastic scintillator coupled to a silicon photomultiplier for positron detection. 
The 30~keV $^{32}$Ar$^+$ ions from ISOLDE are implanted on an about $7~\mu m$ thick $\phi=15$~mm mylar catcher at the center of the setup. Positrons emitted in the upper hemisphere are confined by a 4~T vertical magnetic field and guided towards the plastic scintillator with an efficiency close to 100\%. 
For protons, the total detection efficiency is about 8\% due to the solid angle. The four upper silicon detectors, labeled Si1U to Si4U, are located 65.5~mm above the catcher and the four lower ones, labeled Si1D to Si4D, are mounted in a mirrored configuration below the catcher.
For protons of a few MeV, the energy resolution of the detectors ranges from 25~keV to 45~keV (FWHM).
All detectors, including the scintillator, were read out by the FASTER data acquisition system \cite{FASTER}. 
During an effective beamtime of 35 hours, $\sim10^5$ proton-positron coincidences were collected for the superallowed 0$^+$~$\rightarrow$~0$^+$ transition, which corresponds to an implantation rate of $\sim 100$~pps. 
Ion transmission in the beamline was only about 12\% due to the inadequate existing beam optics. $^{32}$Ar$^+$ ions were produced by a 1.4~GeV proton beam with a mean intensity of 1.4~$\mu$A driven by the CERN Proton Synchrotron Booster and impinging on a CaO target. Ions extracted from the VADIS (Versatile Arc Discharge Ion Source) ion source were then mass selected using the ISOLDE High Resolution mass Separator. The average $^{32}$Ar$^+$ production yield was estimated to be $\approx$1700 pps, more than a factor two below the ISOLDE standard capability \cite{ISOLDE}. With the nominal ion production yield and an improved beam transmission, the present $^{32}$Ar$^+$ implantation rate can thus be increased by more than one order of magnitude in future experiments.
\begin{figure}[h!]
\includegraphics[width=\columnwidth]{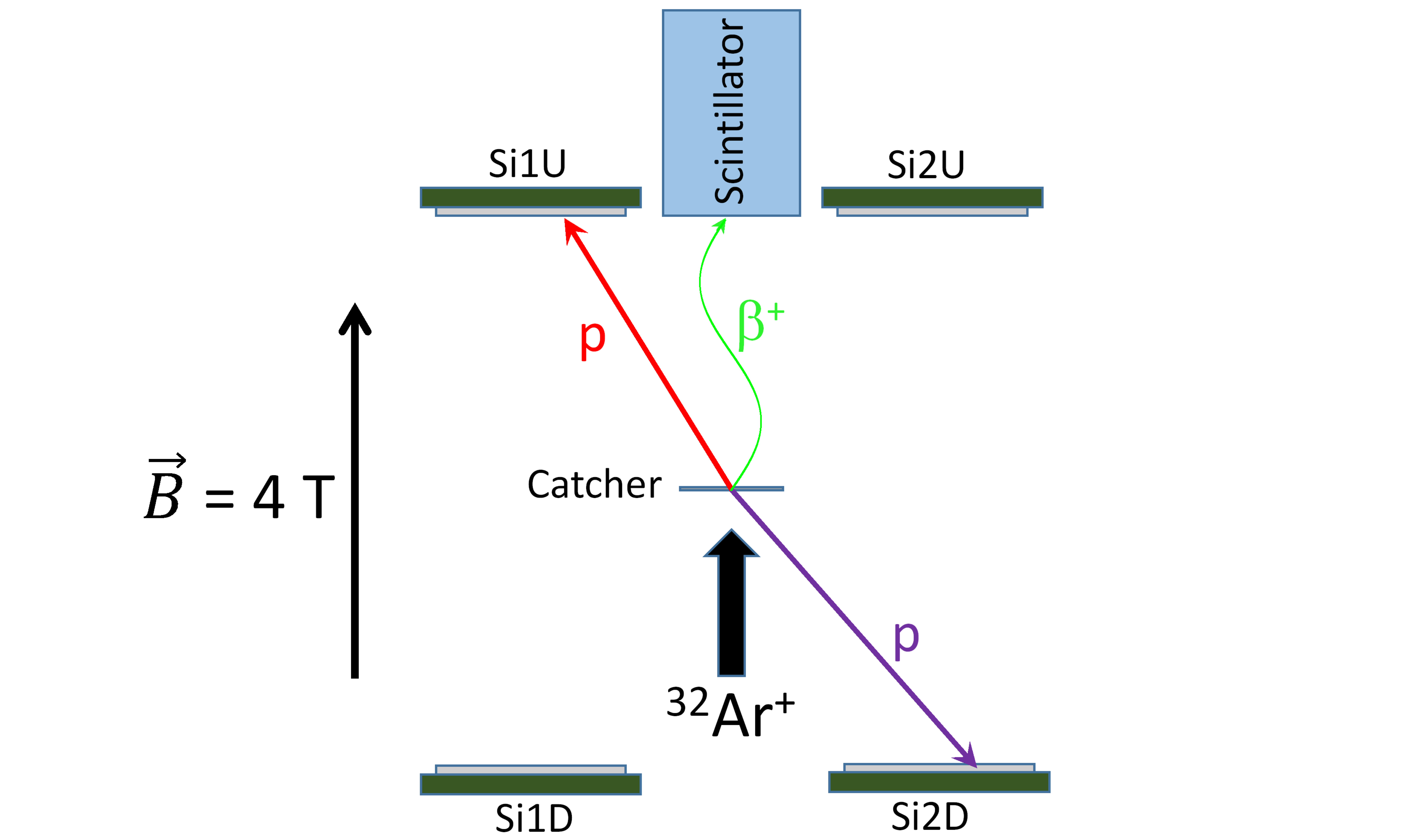}
\caption{(Color online) Schematic of the detection setup (see text for details). Only four silicon detectors are visible on this sectional view. The energy difference between protons emitted in the same hemisphere as the beta particle (red) and those emitted in the opposite one (purple) is a function of ${\tilde a_{\beta\nu}}$.}
\label{fig2}
\end{figure}

\section{\label{sec3}DATA ANALYSIS}
The silicon detectors were calibrated using the six proton peak energies indicated in Fig.~3(a). 
The mean proton energies and their associated uncertainties were previously inferred from an independent experiment performed at GANIL with a detector calibration based on five accurate transition energies in $^{33}$Ar beta-delayed proton decay \cite{Adimi2010}.
In the present experiment, the calibration of the detectors accounts for proton energy loss in both the silicon detector dead layer and the catcher. Mean energy losses were determined using the $\it pstar$ NIST tables \cite{pstar}. For the four lower detectors, proton peaks are not shifted by energy loss in the catcher. The detectors dead layer contribution alone could thus be inferred from a global fit assuming linear calibration functions and an identical dead layer thickness for all detectors. 
The resulting value for the dead layer thickness, $\epsilon=430(300)$~nm, comes with a rather large uncertainty due to the fitting procedure. Note that this uncertainty will be strongly reduced in future experiments by using dedicated detectors with both a thinner and well-known dead layer.
The detector dead layer causes the small energy shifts (a few keV) of the proton peaks with respect to the mean incident energies indicated by dashed lines in Fig.~3(a).
For the upper detectors (Fig.~3(b)), an additional energy loss in the 6.70(15)~$\mu$m thick mylar catcher foil was accounted for. The mylar thickness was inferred from the energy differences between the main proton peaks, shifted towards lower energy, and weaker background peaks visible at the same positions as in Fig.~3(a) for 2123~keV and 3356~keV protons. 
These background peaks result from protons emitted directly towards the upper detectors by $^{32}$Ar ions implanted on the walls and on the setup structure or by outgassing $^{32}$Ar atoms.
For all detectors, the relative uncertainty on the calibration slope was found to be 0.2$\%$ and is dominated by the detectors dead layer contribution.
%
\begin{figure}[h!]
\includegraphics[width=\columnwidth]{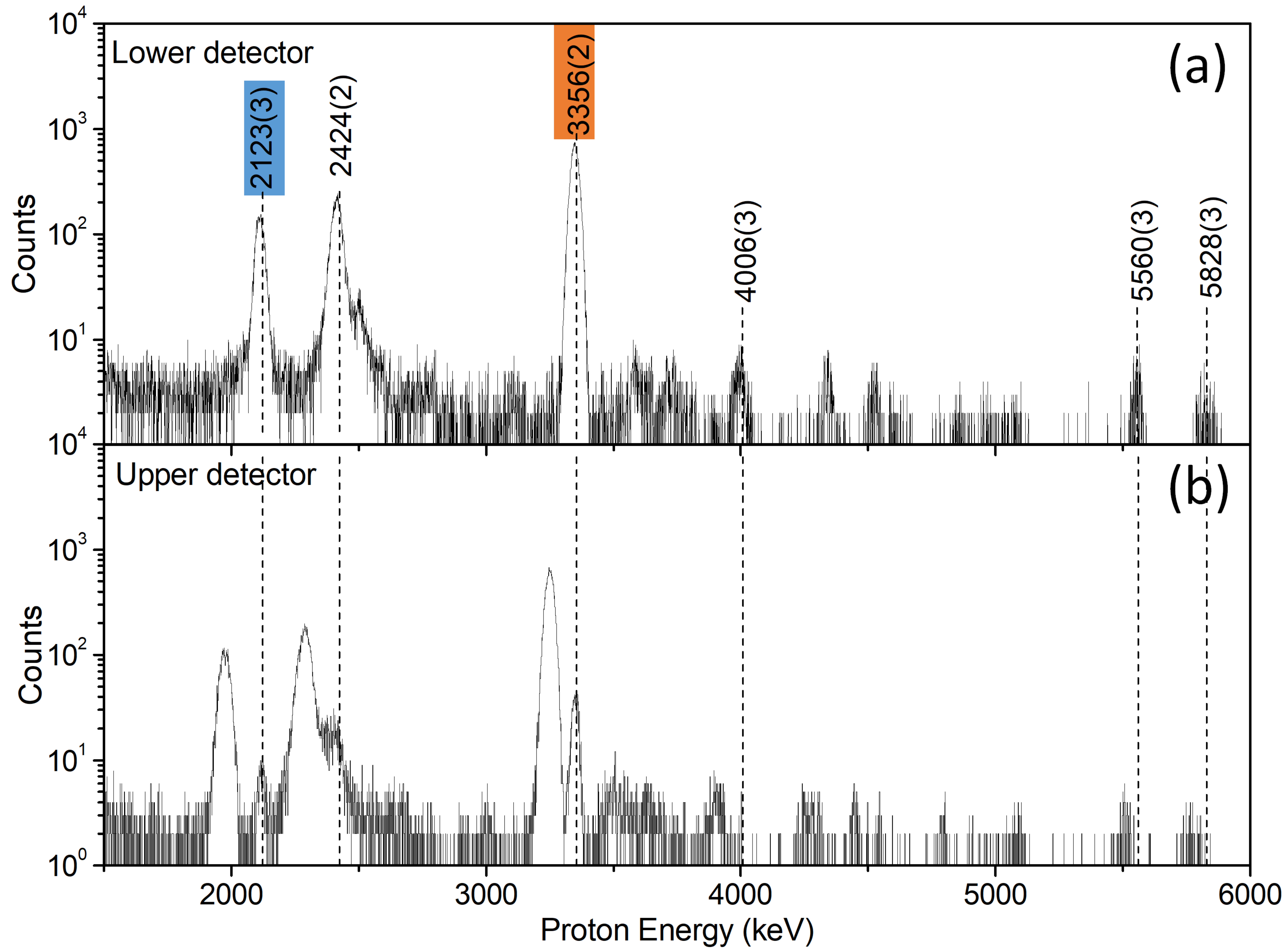}
\caption{(Color online) Deposited proton energy for one lower (a) and one upper (b) silicon detector. The mean proton energies in the laboratory of the six proton peaks used for the calibration are indicated by dashed lines. Proton energies corresponding to the F and GT transitions of interest are highlighted in orange and blue, respectively.}
\label{fig3}
\end{figure}
%

The pure F and GT transitions leading to 3356~keV and 2123~keV protons, respectively, were both studied with the same analysis method.
Two sets of events were selected: one with events defined as singles, {\it i.e.} without condition on the beta particle, and one with only coincidences, where both a positron and a proton are detected. For each detector, singles provide a mean proton energy reference independent of ${\tilde a_{\beta\nu}}$ as only the shape and width of the peaks are affected by the daughter recoil. The mean proton energy shift between coincidences and singles for upper and lower detectors is then a function of the value of ${\tilde a_{\beta\nu}}$, of the beta energy distribution, and of the beta-proton angular distribution. 
The selection of coincidences for the F transition is illustrated in Figs.~4(a) and 4(b) where a cut was performed on the proton energy and on the trigger time difference $T_{\mbox{\scriptsize{diff}}}=T_p-T_{ \beta}$, where $T_p$ and $T_{ \beta}$ are the trigger times of the proton and positron detectors. The time of flight of the protons is of the order of $\sim3$~ns and the $T_{\mbox{\scriptsize{diff}}}$ distribution results here primarily from the detector response functions. The proton energy cut shown in Fig.~4(a) is a 100~keV window centered on the mean energy of the superallowed delayed-proton peak and the $T_{\mbox{\scriptsize{diff}}}$ selection of Fig.~4(b) was set between 50~ns and 350~ns. Events with $T_{\mbox{\scriptsize{diff}}}$ outside the time selection criteria were used to estimate the contribution of false coincidences which turned out to be less than 0.3$\%$ for the selected data. Their effect on the energy shift and the associated error on the correlation measurement (see Table I) were determined by varying the width of the $T_{\mbox{\scriptsize{diff}}}$ window up to 1000~ns.
%
\begin{figure}[h!]
\includegraphics[width=\columnwidth]{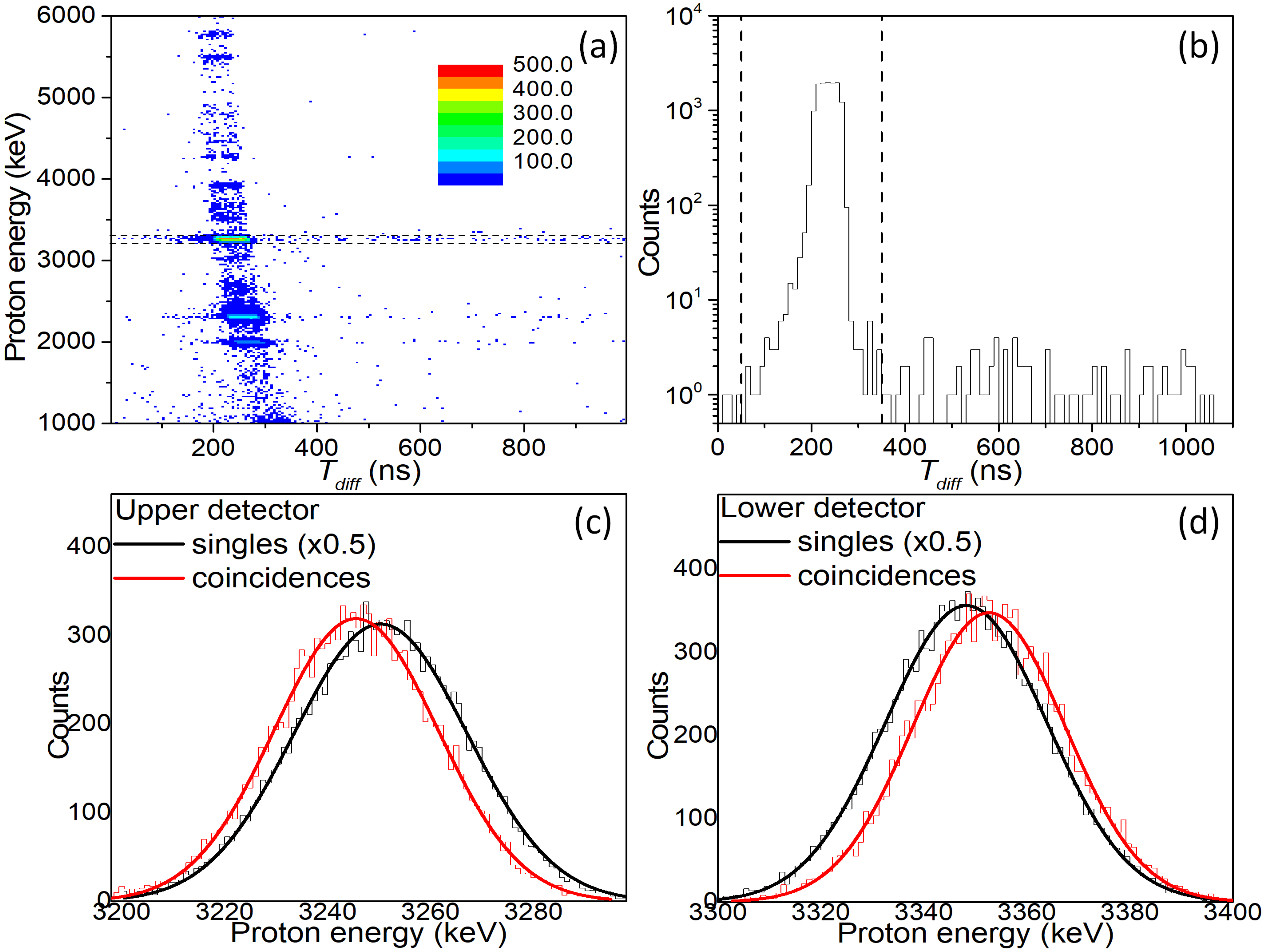}
\caption{(Color online) (a): Proton energy spectrum versus $T_{\mbox{\scriptsize{diff}}}$ for one of the upper detectors. The energy selection window is indicated by dashed lines. (b): $T_{\mbox{\scriptsize{diff}}}$ spectrum for the selected events of Fig. 4(a). The coincidence selection window is given by the vertical dashed lines. (c) and (d): Proton energy of the superallowed transition for coincidences (red) and singles (black) obtained for one upper and one lower detector. The Gaussian fits are to guide the eye.}
\label{fig4}
\end{figure}

Figs. 4(c) and 4(d) show the resulting proton energy peaks for singles and coincidences obtained for the F transition for one of the upper and one of the lower detectors. 
As only positrons emitted in the upper hemisphere can be detected, the data obtained for singles was scaled by a factor 0.5 on both figures to ease the comparison with the coincidences.
The energy shifts of coincidences towards lower values for the upper detector and towards higher values for the lower detector are clearly observed. Peaks obtained for singles are also wider, due to the larger kinetic energy broadening arising when all emission angles are allowed for the beta particle. 
For each peak, the mean energies $\overline{E}_{coinc}$ (for coincidences) and $\overline{E}_{single}$ (for singles) were determined using an iterative procedure by selecting events within a 100~keV window centered on the mean energy obtained in the previous step. This method was favored compared to the use of fits as it does not require any knowledge of the exact detector response function and does not depend on the shape of the intrinsic proton energy distribution in the center of mass. To limit the influence of calibration imperfections, the mean energy shifts of coincidences were determined individually for each detector using data from singles as a reference. Since the energy loss in the detector dead layer and in the catcher depends on the incident proton energy, it is slightly different for coincidences and singles. Measured energy shifts for the 3356~keV protons of the F transition were thus all corrected from a small increase of ${\delta}E_{lower}$=0.007(5)~keV (for lower detectors) and of ${\delta}E_{upper}$=0.099(7)~keV (for upper detectors) determined using the $\it{pstar}$ NIST data base. 
\begin{figure}[h!]
\includegraphics[width=\columnwidth]{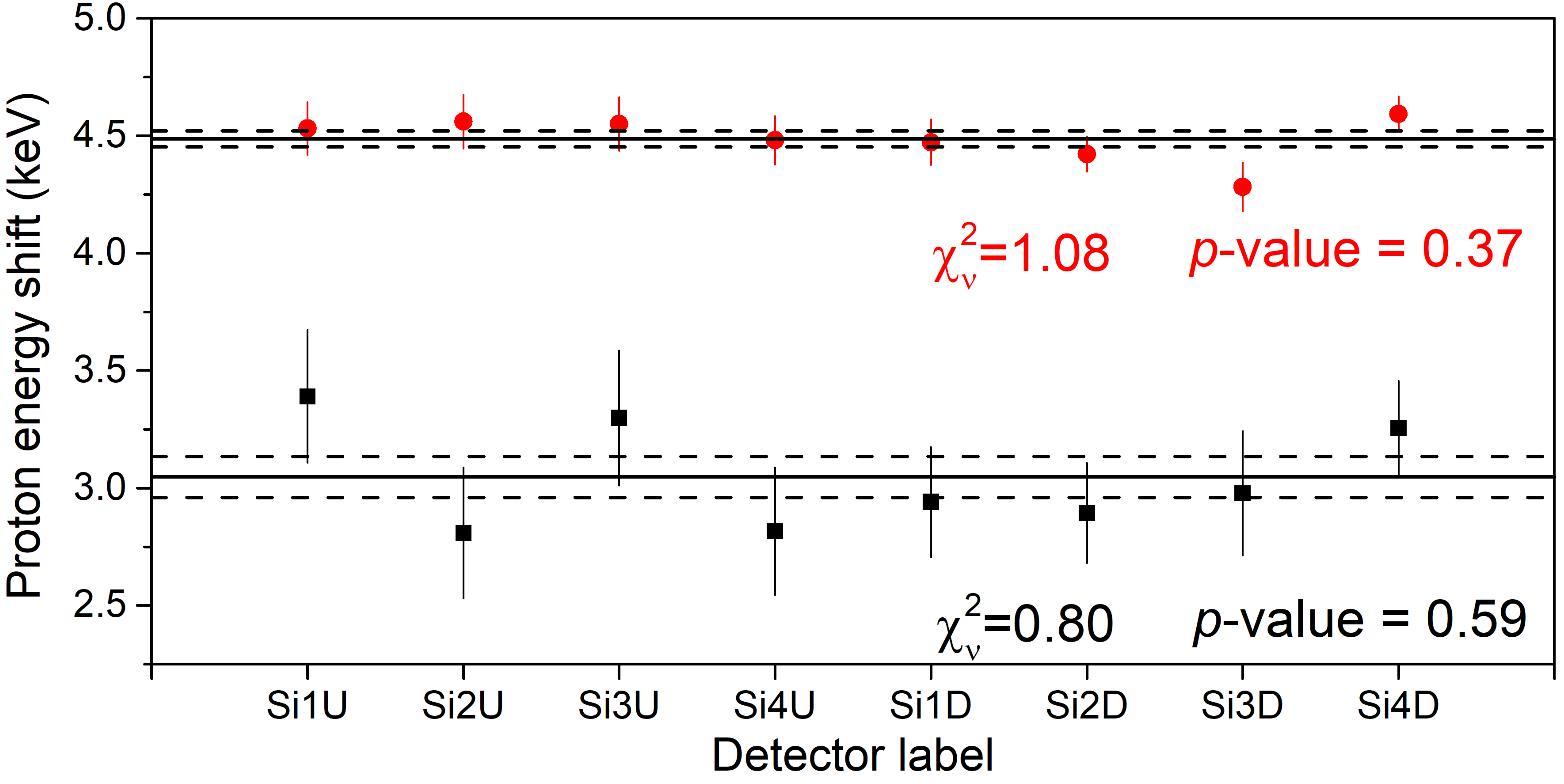}
\caption{(Color online) Proton mean kinetic energy shifts (absolute values) and associated statistical errors obtained with the upper and lower detectors for the 3356~keV protons from F decay (red circles) and the 2123~keV protons from GT decay (black squares). Weighted averages and statistical uncertainties are indicated by solid and dashed lines, respectively.} 
\label{fig5}
\end{figure}

The mean kinetic energy shifts $\overline{E}_{shift}={\mid}\overline{E}_{coinc}-\overline{E}_{single}{\mid}-{\delta}E_{upper/lower}$ obtained for all detectors in the superallowed transition are given in Fig. 5 (red circles), with error bars corresponding to statistical uncertainties only. The same analysis procedure was also applied to the set of events from one of the pure GT transitions of the $^{32}$Ar decay with 2123~keV proton emission (black squares). 
The resulting weighted average energy shifts are 4.49(3)~keV and 3.05(9)~keV for the F transition and for the GT transition, respectively, with associated reduced $\chi^2$ and $p$-values showing a good compatibility of the results obtained with eight different detectors.

\section{\label{sec4}SIMULATIONS}
Despite the relative simplicity of the setup, the relationship between the mean kinetic energy shift $\overline{E}_{shift}$ and the value of ${\tilde a_{\beta\nu}}$ can only be precisely established using Monte Carlo simulations. The decay kinematics was simulated using the decay rate given by Eq.~(3) and for a maximum positron kinetic energy of 5065(2)~keV for the F transition and of 6339(2)~keV for the GT transition. 
Beside the Fermi function, theoretical corrections described in Ref.~\cite{Hayen18} that may contribute up to the $10^{-2}$ level were not yet included. The theoretical uncertainty due to such corrections will remain below $10^{-3}$.
 In the simulations, $^{32}$Ar decay sources were set on the lower surface of the catcher foil using several spatial distributions. Delayed protons were emitted randomly in all directions by the $^{32}$Cl daughter nuclei, taking into account the recoil-induced energy shift. Proton trajectories within the $B=4$~T magnetic field of the setup were computed analytically, considering only energy losses in the catcher and detectors dead layer.
Using the TRIM simulation toolkit \cite{TRIM}, proton straggling and backscattering in the catcher and in the detectors dead layer were investigated on beforehand and found to be negligible. Mean proton energy losses in dead layers obtained by using the $\it pstar$ NIST tables and TRIM were also compared and found to agree within less than 2.5\%.
To study systematic effects associated with proton detection, the backscattering of the beta particles in the catcher and in the plastic detector were in a first step fully neglected and all positrons emitted in the upper hemisphere were considered as detected. Simulations with $10^7$ coincidences were ran with five values of ${\tilde a_{\beta\nu}}$ ranging from -1 to 1 and considering $b=0$.
The relationship between ${\tilde a_{\beta\nu}}$ and the proton energy shift $E_{shift}$ was found to be perfectly linear for both transitions with a slope $d\tilde a_{\beta\nu}/dE_{shift}=0.9684(2)$~ keV$^{-1}$ for the Fermi transition and $d\tilde a_{\beta\nu}/dE_{shift}=0.9788(2)$~ keV$^{-1}$ for the GT one.
Systematic errors associated to imperfections of the setup such as the $B$ field strength in both upper and lower sections, the positions of the silicon detectors and the distribution of the implanted ions on the catcher were  estimated  to first order approximation by scanning each parameter individually. The sensitivity of the measurement to these parameters and associated errors are summarized in Table~I. The dominant contributions for proton detection are due to the uncertainty on the detector calibration slopes and on the dead layer thickness.

\begin{table}[!htb]
\caption{Sources of systematic error, uncertainties on the source of error and associated uncertainties on ${\tilde a_{\beta\nu}} (\times 10^{-3})$ for the F and GT transitions. Two additional digits (not displayed here) were accounted for in the calculation of the quadratic sum provided as total systematic error.}
\label{tab:results}
\begin{ruledtabular}
\begin{tabular}{ccccc}
 & Source & Uncertainty & ${\Delta \tilde a_{F}}$ & ${\Delta \tilde a_{GT}}$\\
\hline
 background & false coinc. & $8\%$ & $<$ 1 & 2\\
\hline
proton & det. calibration & 0.2\% & 9 & 6\\
 & det. position & 1 mm & $<$ 1 & 1\\
 & source position & 3 mm & 3 & 2\\
 & source radius & 3 mm & 1 & 1\\
 & B field & 1\% & $<$ 1 & $<$ 1\\
 & silicon dead layer & 0.3 $\mu$m & 5 & 7\\
 & mylar thickness & 0.15 $\mu$m & 2 & 3\\
\hline
positron & detector backscattering & 15\% & 2 & 1\\
 & catcher backscattering & 15\% & 21 & 11\\
 & threshold & 12 keV & 8 & 4\\
\hline
total & & & 25 & 16\\
\end{tabular}
\end{ruledtabular}
\end{table}

In a second step, the simulation package GEANT4 \cite{GEANT4} was used to account for positron backscattering in the catcher foil and in the plastic scintillator.
Figure 6(a) shows for the F transition the deposited energy spectra in the scintillator obtained experimentally and by simulations with or without positron scattering. The detection threshold indicated in Fig. 6(b) was estimated to be 25(12)~keV, with a conservative 12~keV uncertainty due to low statistics and to a limited knowledge of the detector response function.
As positrons are confined by the magnetic field, some of their trajectories have grazing incidences when reaching the plastic scintillator, which increases backscattering from the detector with a reduced deposited energy. The simulation shows that $\sim36\%$ of the positrons escape from the detector volume with a remaining kinetic energy larger than 50~keV. This strong yield of positron backscattering and the shortcomings of GEANT4 lead to discrepancies between simulated (red) and experimental (black) data at the level of 10-15\%. We thus considered a 15\% relative uncertainty on the backscattering correction provided by GEANT4, which is larger than the typical discrepancies obtained in previous studies (see Ref.~\cite{Soti13, Dondero18} and references therein).
\begin{figure}[h!]
\includegraphics[width=\columnwidth]{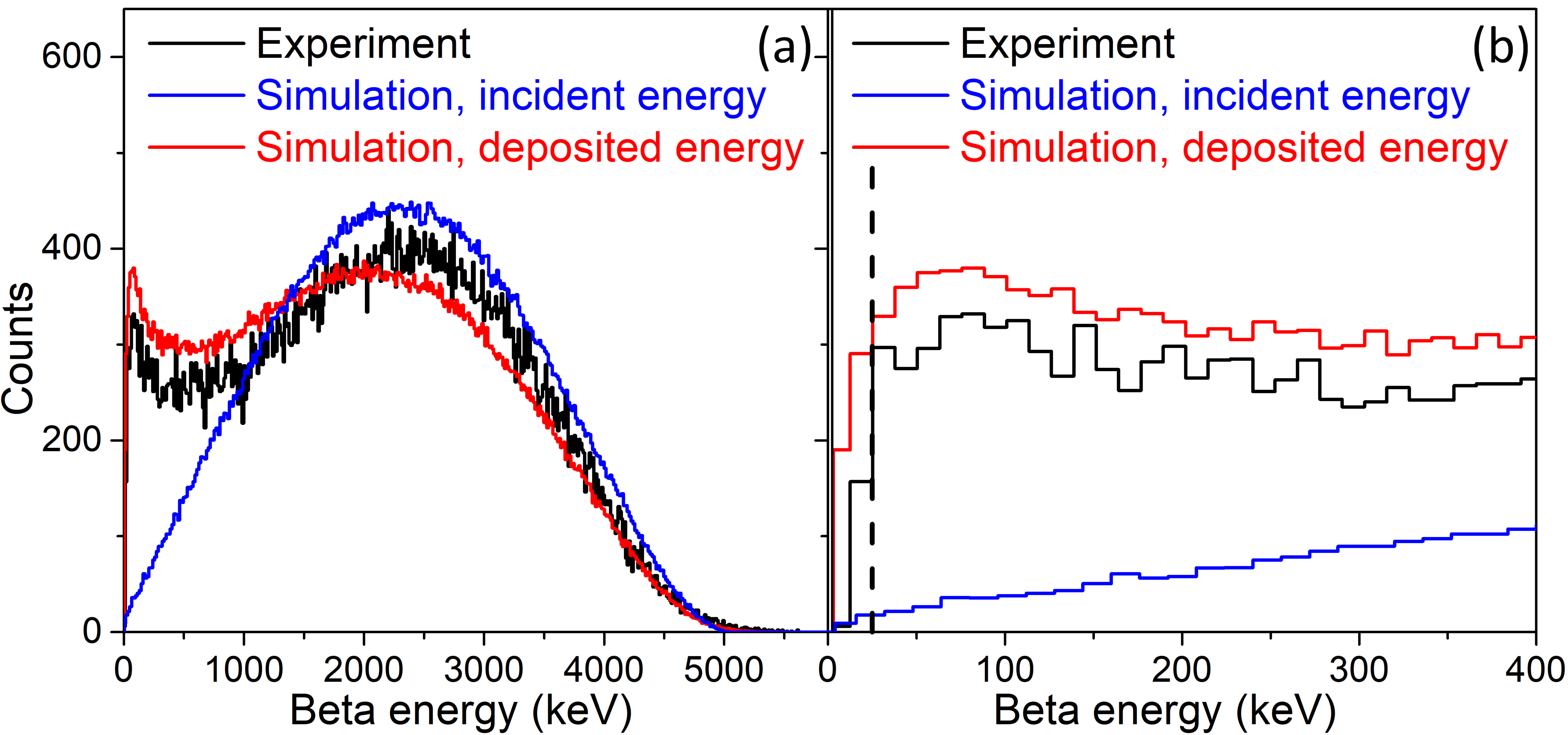}
\caption{(Color online) (a): Incident (blue line) and deposited (red line) positron energy distributions given by the simulation compared to experimental data (black line) for the F transition. (b): Same as (a) with a zoom on the low energy part. The 25~keV threshold is indicated by a dashed line.}
\label{fig6}
\end{figure}

We must stress that the measurement of the energy of the positrons does not play a direct role in the extraction of ${\tilde a_{\beta\nu}}$ which is inferred from the proton peak energy shift alone. The contribution of positron backscattering arises only from the fact that this process can lead to the non-detection of a fraction of the positrons initially emitted in the upward direction. Such events are counted as singles in the experimental data and must be accounted for in the simulation.Thanks to a very low detection threshold (25~keV for a maximum positron energy of more than 5~MeV), the fraction of positrons that are not detected due to backscattering on the detector was found to be $3.1\times 10^{-3}$ for the F transition and $2.5\times 10^{-3}$ for the GT one. The relative error of 15\% on these values which were obtained using GEANT4 leads to an uncertainty on $\tilde a_{\beta\nu}$ of $2.3\times 10^{-3}$ and $1.1\times 10^{-3}$ for the F and GT transitions respectively, as provided in Table~I.
The systematic error due to the limited knowledge of the detection threshold is about four times larger. Positron backscattering on the catcher is another source of systematic error as it prevents the detection of positrons initially emitted upwards. For the F transition, a backscattering rate from the catcher of 3.3(5)\% was obtained using GEANT4. The 0.5\% uncertainty on this rate arises from the 15\% relative uncertainty on the GEANT4 backscattering correction. The associated uncertainties on $\tilde a_{\beta\nu}$ for the F and GT transitions in Table I are the dominant ones. This loss due to backscattering in the catcher can also be independently estimated from the experimental data by comparing the number of coincidences and of singles detected in the proton upper detectors. This experimental estimate was found to be 3.5(4)\%, in perfect agreement with the GEANT4 simulation. With higher statistics, this measurement of the experimental backscattering rate will be a strong asset to validate the accuracy of GEANT4.
\section{\label{sec5}RESULTS AND DISCUSSION}
The results of this proof-of-principle experiment based on proton-positron coincidences are ${\tilde a_{\beta\nu}=1.007(32)_{stat}(25)_{syst}}$ for the Fermi transition and ${\tilde a_{\beta\nu}=-0.222(86)_{stat}(16)_{syst}}$ for the GT transition. Both values are found in agreement with the SM predictions with deviations of $0.2\times \sigma$ and $1.3\times\sigma$ for the F and GT transitions respectively when considering statistical errors only. Despite the use of a rudimentary setup and a very short beam time allocated for this test, the collected data provides the third most precise measurement of ${\tilde a_{\beta\nu}}$ in a pure F transition.

The sensitivity to the Fierz interference term, $b$, was also determined by assuming only left-handed neutrinos and by running simulations with different values of $b$ ranging from $-0.1$ to $+0.1$ (the value of ${ a_{\beta\nu}}$ being modified accordingly). This sensitivity to $b$ is characterized by the coefficient $\alpha$ mentioned in the introduction. We found $\alpha \simeq 0.35$ and $\alpha \simeq -0.11$ for the F and GT transitions, respectively. For the F transition, this is a gain in sensitivity on $b_{F}$ of a factor $\simeq1.8$ when compared to the value $\alpha \simeq 0.19$ obtained in Ref. \cite{Adelberger99}. Thanks to this higher sensitivity, the constraints on scalar couplings for left-handed neutrinos provided by this test experiment are in fact only a factor of 3.5 less stringent than the ones inferred from Ref. \cite{Adelberger99}.
Oppositely, the sensitivity on $b_{GT}$ is very low: assuming the same uncertainty on ${\tilde a_{F}}$ and ${\tilde a_{GT}}$ in our experiment, the corresponding uncertainty on $b_{GT}$ would be a factor $\sim 10$ larger than on $b_{F}$. The technique is thus not well adapted to search for exotic tensor couplings with left-handed neutrinos. The simultaneous measurement of ${\tilde a_{\beta\nu}}$ for the GT transition with a high precision remains nevertheless very useful to check systematic errors and the validity of the analysis. The constraints on $b_{GT}$ provided in Table 8 of Ref. \cite{Gonzalez19}, with an uncertainty of $3.9\times 10^{-3}$, correspond to an uncertainty of $1.4\times 10^{-4}$ on ${\tilde a_{GT}}$ with the present setup. A measurement of ${\tilde a_{GT}}$ for a GT transition must thus be in agreement with the SM at this level of precision of $\sim 10^{-4}$. As the sources of systematic errors are similar for both the F and the GT transitions, such an agreement will be mandatory to validate the results obtained with the F transition.

\section{\label{sec6}PROSPECTS}
A new measurement is planned at ISOLDE/CERN soon after the restart of the CERN accelerators. Within a two-week beam time, assuming nominal $^{32}$Ar production from the ISOLDE target, a beam transmission increased from 12\% to 70\% and a detection solid angle for protons increased by a factor of three, the statistical error will be reduced by more than a factor 20. The segmented silicon detector designed for this future measurement will not only provide a three times higher detection solid angle but also a higher resolution (between 5~keV and 10~keV) and a thinner dead layer (80(20)~nm) \cite{Atanasov}. With a resolution of 10~keV on the proton energy, which is rather conservative, the total gain factor expected on the statistical error is close to 50. In addition, the improved resolution will allow to measure ${\tilde a_{GT}}$ for the GT transition leading to the intense proton peak at 2424~keV which was not fully resolved with the present setup. Under these conditions, the expected statistical errors on ${\tilde a_{F}}$ and ${\tilde a_{GT}}$ are $0.7\times 10^{-3}$ and $1.1\times 10^{-3}$, respectively.

Reducing the systematic error at the $10^{-3}$ level will be more challenging, but the present efforts are focused on this goal. The new setup under development will ensure that all uncertainties of Table I that are due to a limited knowledge of the detection system (detector dead layer, source and detector relative positions, magnetic field homogeneity) will be reduced by a factor $\sim 5$. For the catcher, we will use commercial 500~nm thick mylar foils that will readily reduce by a factor $\sim 13$ the effect of positron backscattering. We plan to alternate measurements with implantation on a 500~nm foil and on a 10~$\mu$m foil serving as a catcher to estimate directly (without relying on simulations) how the foil thickness impacts the extraction of ${\tilde a_{\beta\nu}}$. The resulting extrapolation will allow to limit the catcher contribution at the level of the statistical error. Off line measurements of the backscattering rate using a $^{90}$Sr source covered by different mylar thicknesses in place of the $^{32}$Ar will be performed and compared to GEANT4 simulations. They should provide an even higher precision on the catcher contribution. A low intensity electron beam is also under development to characterize the response function of the positron detector in the 0-30 keV range. The goal is to lower the positron energy threshold below 10~keV and reduce the uncertainty on this threshold down to 1~keV. The systematic error on ${ \tilde a_{\beta\nu}}$ directly due to the threshold will then be reduced down to $0.7\times 10^{-3}$. Lowering the threshold below 10~keV will allow to reduce the contribution of positron backscattering on the detector down to $0.9\times 10^{-3}$. In parallel, dedicated measurements are ongoing with an electron spectrometer to characterize more precisely beta particle backscattering in the scintillator and further reduce this source of uncertainty. Last but not least, the uncertainty due to the proton detector calibration slope must be improved by one order of magnitude to reach the $10^{-3}$ precision level. As the dead layer of the future proton detectors will have a negligible contribution to the calibration uncertainty, an improvement by a factor of $\sim 5$ seems readily achievable by using the $^{33}$Ar proton peaks energy provided in Ref. \cite{Adimi2010} as a reference. 
\begin{figure}[h!]
\includegraphics[width=\columnwidth]{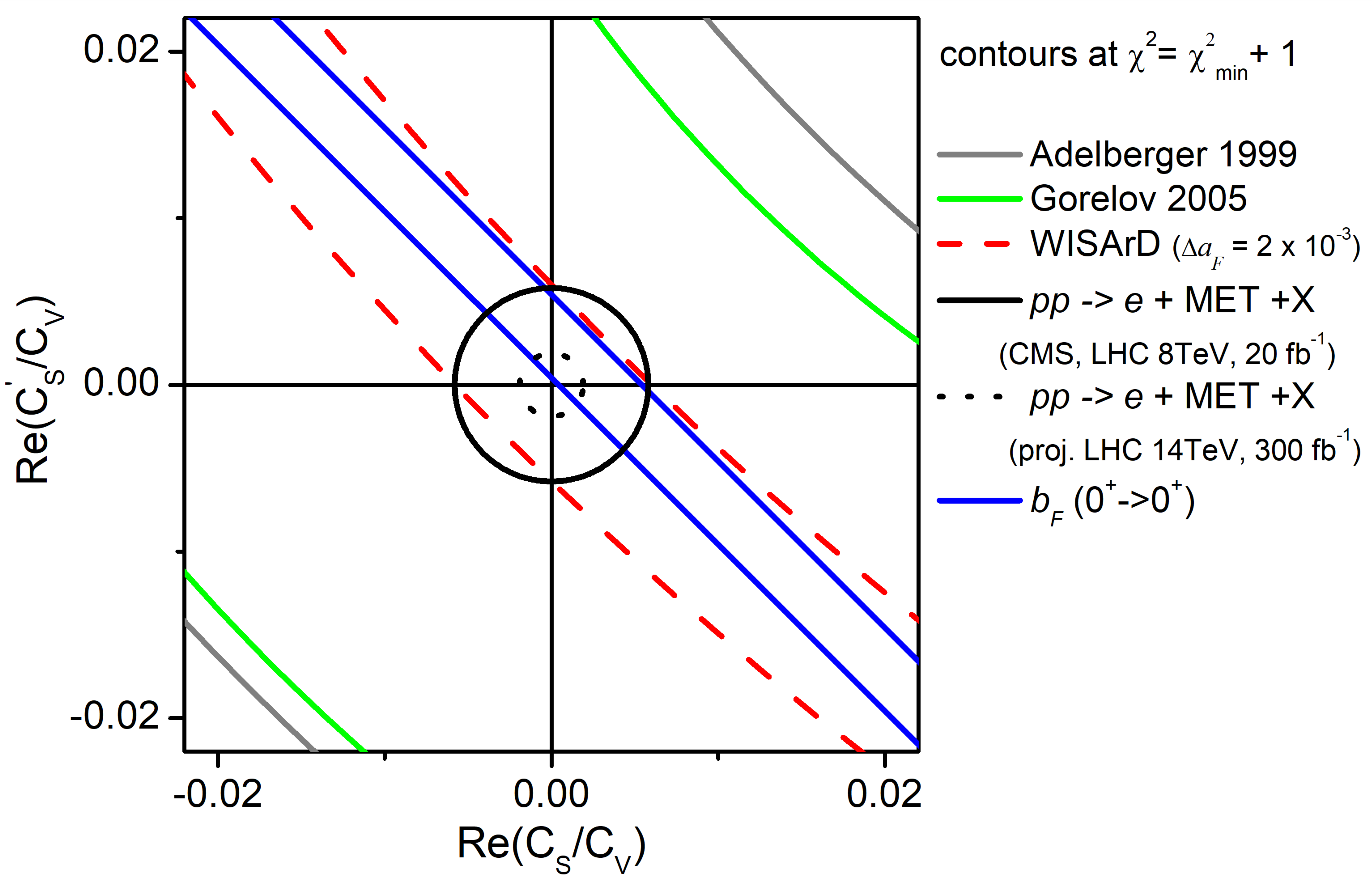}
\caption{(Color online) Present (plain lines) and expected (dotted lines) constraints on scalar coupling constants extracted from Ref. \cite{Gonzalez19} (and references there in). Contours labeled 'Gorelov 2005' are calculated using the values of ${ \tilde a_{\beta\nu}}$ and $\alpha$ provided in Ref. \cite{Gorelov2008}. The expected constraints for the next WISArD measurement with a $2\times 10^{-3}$ precision are indicated by dashed red lines. All exclusion contours are given for one standard deviation.}
\label{fig7}
\end{figure}

According to these projections, the next measurement of ${\tilde a_{\beta\nu}}$ in the pure F transition of $^{32}$Ar could reach a precision of the order of $2\times 10^{-3}$ with an uncertainty dominated by the contribution of the proton detector calibration slope. The corresponding constraints on scalar couplings are shown in Fig.~7 along with those from Ref. \cite{Gonzalez19}, including present and future search at the LHC, constraints provided by $Ft$ value measurements in superallowed transitions and previous correlation measurements. With this level of precision and thanks to a higher sensitivity to the Fierz interference term, the future measurement will improve significantly the present constraints on exotic currents of the weak interaction inferred from correlation measurements in nuclear beta decay and remain competitive with the search at the LHC.
Further improvements will be pursued after the next experimental campaign with the aim to lower the systematic uncertainty below $10^{-3}$.
 \\
\section*{ACKNOWLEDGMENTS}
We thank the ISOLDE staff for their technical support during the experiment and A. Garc\'ia for fruitful discussions. This work was partly funded by the ANR-18-CE31-0004-02, by the European Union's 7th Framework Programme, Contract No. 262010 (ENSAR), by the Flemish Fund for Scientific Research FWO, and by the Grant of the Ministry of Education of the Czech Republic LTT17018.


\end{document}